# Effect of passengers' active head tilt and opening/closure of eyes on motion sickness in lateral acceleration environment of cars

Takahiro Wada and Keigo Yoshida

*Takahiro Wada*

*College of Information Science and Engineering, Ritsumeikan University, 1-1-1 Noji-higashi, Kusatsu, Shiga 525-8577, Japan*

*Email: twada@fc.ritsumei.ac.jp*

*Tel.: +81-77-561-2798*

*Keigo Yoshida\**

*College of Information Science and Engineering, Ritsumeikan University, 1-1-1 Noji-higashi, Kusatsu, Shiga 525-8577, Japan*

*Email: keigo4869@gmail.com*

*Tel.: +81-77-561-2798*

*\*Mr. Yoshida moved to Omron Automotive Electronics Co. Ltd. after completing this research.*

Corresponding author:

Takahiro Wada, College of Information Science and Engineering, Ritsumeikan University

**Acknowledgments**

The authors are grateful to Yoshimitsu Kawai, Makoto Sumikawa, and Ena Dejima of Ritsumeikan University for their assistance in data collection. The authors would also like to thank Tsukinowa Driving School for allowing us to use their driving course.

**Word count:** 5,826 words

# Effect of passengers' active head tilt and opening/closure of eyes on motion sickness in lateral acceleration environment of cars

This study examined the effect of passengers' active head-tilt and eyes-open/closed conditions on the severity of motion sickness in the lateral acceleration environment of cars. In the centrifugal head-tilt condition, participants intentionally tilted their heads towards the centrifugal force, whereas in the centripetal head-tilt condition, the participants tilted their heads against the centrifugal acceleration. The eyes-open and eyes-closed cases were investigated for each head-tilt condition. In the experimental runs, the sickness rating in the centripetal head-tilt condition was significantly lower than that in the centrifugal head-tilt condition. Moreover, the sickness rating in the eyes-open condition was significantly lower than that in the eyes-closed condition. The results suggest that an active head-tilt motion against the centrifugal acceleration reduces the severity of motion sickness both in the eyes-open and eyes-closed conditions. They also demonstrate that the eyes-open condition significantly reduces the motion sickness even when the head-tilt strategy is used.

**Practitioner Summary:**

Little is known about the effect of head-tilt strategies on motion sickness. This study investigated the effects of head-tilt direction and eyes-open/closed conditions on motion sickness during slalom automobile driving. Passengers' active head tilt towards the centripetal direction and the eyes-open condition greatly reduce the severity of motion sickness.

**Introduction**

Many experimental studies on human exposure to whole-body vibration have investigated the susceptibility of humans to motion sickness at many frequencies and amplitudes of vibration (Griffin 1990). These studies have revealed that motion sickness is most acute for linear vibrations of around 0.1–0.3 Hz in the vertical (O'Hanlon and McCauley 1974), fore–aft (Golding, Mueller, and Gresty 2001), and lateral (Griffin and Mills 2002) directions and that the severity of motion sickness increases with the magnitude of vibration. It has also been shown that linear vibrations greater than 1 Hz hardly provoke motion sickness even though they cause discomfort (Griffin 1990). Moreover, investigations of passive whole-body vibrations in rotational motion (Howarth and Griffin 2003) and a combination of linear and rotational motions (Joseph and Griffin 2007; Butler and Griffin 2009) have revealed the complicated features of motion sickness susceptibility.

Regarding carsickness, it has been shown that automobile vibrations in the vertical direction do not generally include vibrations in the most provocative frequency range around 0.2 Hz (Griffin and Newman 2004) and therefore do not correlate with motion sickness (Salvendy 2006; Griffin 1990; Kato and Kitazaki 2006; Corbridge, Griffin, and Whitham 1986). On the other hand, Vogel, Kohlhaas, and von Baumgarten (1982) reported that repeated hard braking caused motion sickness. Kato and Kitazaki (2006) measured the severity of motion sickness in 87 participants when riding different mini-vans and analysed the cause of the sickness using an expanded version of the motion sickness dose value (ISO 2631-1, 1997) for three-dimensional motion. Their study revealed that the severity of motion sickness is correlated with the fore–aft and lateral accelerations of the head motion but not with the vertical motion. In addition, Griffin and Newman (2004) deduced that motion sickness was more severe when the participants could not see the upcoming road through the front window of the automobile.

Some studies have reported that the riding conditions of the passengers also affect the severity of motion sickness. For example, Rolnick and Lubow (1991) investigated why the automobile driver is generally less susceptible to carsickness than the passengers and observed that participants who controlled the vehicle were less likely to experience motion sickness than those without control, who were 'yoked' mechanically to the vehicle. Wada et al. (2012) investigated the effect of the head tilt of individuals seated in the navigator seat of a passenger car on the severity of motion sickness. The motivation for their research was based on the fact that the driver is known to tilt his or her head towards the curve centre when negotiating a curve, whereas the passenger's head moves in the opposite direction (Fukuda 1976; Zikovitz and Harris 1999; Wada, Kamiji, and Doi 2013). Their study (Wada et al., 2012) revealed that the passengers' head tilt against the centrifugal force in slalom driving, which imitated the driver's head movement, significantly reduced the severity of motion sickness compared with the natural head tilt towards the centrifugal force. However, all the participants had their eyes open during these experiments, and no studies have investigated how the open/closed condition of the eyes affects the motion sickness with the active head-tilt strategy. In addition, the magnitudes of the head movements of the participants were not uniform in the study by Wada et al. (2012) because the head tilted actively in the centripetal direction but naturally (passively) in the centrifugal direction.

By expanding the results of the one-degree-of-freedom (1DOF) model developed by Bos and Bles (1998), Kamiji et al. (2007) proposed a 6DOF mathematical model for the subjective vertical conflict (SVC) theory of motion sickness (Bles et al., 1998), which supposes that motion sickness is caused by the accumulation of conflict between the sensed and estimated vertical directions. The 6DOF mathematical model is composed of transfer functions of the semi-circular canal, otolith, and their interaction as well as their internal model, which is thought to be built in the central nervous system. Motion sickness incidence is estimated as the model output obtained from the model inputs, i.e. the head acceleration and its angular velocity. Using this 6DOF model, Wada, Kamiji, and Doi (2013) predicted that head movement towards the centripetal direction would reduce motion sickness incidence. However, their model did not consider visual information.

Based on these findings, we hypothesised that passengers' head movement against the centrifugal direction reduces motion sickness regardless of whether their eyes are open or closed. We also hypothesised that adding visual information reduces motion sickness because such information is thought to increase the accuracy of the sensed vertical direction. The purpose of the present study is to examine these hypotheses and investigate the effect of their interactions. We therefore conducted experiments using a real passenger car and investigated the effects of the active head-tilt and eyes-open/closed conditions of passengers on the severity of motion sickness in a lateral acceleration environment.

**Method**

*Design*

The experimental design was based on the work of Wada et al. (2012). Ten participants were exposed to an acceleration stimulus while seated in the navigator seat of a passenger car. Two independent variables were considered in the design: (1) the head-tilt condition and (2) the eyes-open/closed condition. The head-tilt condition was either in the centripetal or in the centrifugal direction. In the centripetal condition, the participants were instructed to tilt their heads actively in the direction against the centrifugal force during slalom driving. In the centrifugal condition, they were instructed to tilt their heads actively in the direction of the centrifugal force. The amplitude of the head tilt in both the conditions was set as 20° so that the head direction gets closer to the gravito-inertial force (GIF) direction at the peak lateral acceleration in the centripetal condition, according to Wada et al. (2012). This head-tilt angle was achieved by practicing the head tilt before the measurement trials. Figure 1 illustrates the typical head postures in the centripetal and centrifugal conditions [Figure 1 near here]. An experimenter sat in the rear seat of the car and informed the participant when each slalom run, which was defined as driving in a zigzag manner through eight pylons on one straight segment, began. In addition, the experimenters were asked to correct the participant's head motion if it was not appropriate, e.g. if there was a large phase delay or the tilt angle was small. However, no such situation was encountered in the experiments.

The eyes-open/closed condition included the eyes-open and eyes-closed cases. In the eyes-open condition, the participant was instructed to look outside through the front window of the vehicle, but no instructions were provided regarding the gaze position. In eyes-closed condition, the participant wore an eye mask throughout the experiment. Both the head-tilt and eyes-open/closed conditions were treated as within-subjects factors. In other words, each participant performed all four combinations of the two variables on four different days, with at least seven days between any two experiments. The sequence of the conditions was randomised to decrease the order effect. In addition, for each participant, the experiments were performed at approximately the same time on the four days. The experiments used two drivers, and each participant was assigned to the same driver on all experiment days.

*Participants*

Ten healthy males, with a mean age of 22.9 years (SD 0.8 years), gave their informed consent to participate in the experiments, which was approved by the Ritsumeikan University Ethics Review Committee for Research Involving Human Participants. The participants were notified that they could become motion sick and vomit because of the experiments and that they could stop the experiment at any time and for any reason. It was found that the experimenters provided incorrect instructions to two participants, so the results for those participants were not analysed. Each participant was paid 16,000 Japanese yen as compensation. Before the experiments, the motion sickness susceptibility of each participant was tested using a revised version of the motion sickness susceptibility questionnaire (Golding 1998). The mean percentile score was 46.9% (SD 22.7%), which represents a wide distribution of susceptibility to motion sickness.

*Method of evaluating motion sickness*

Subjective evaluations and symptom score tests were conducted to analyse the severity of the motion sickness of the participants.

For the subjective evaluations, the subjective sickness rating method, used in the studies by Golding, Markey, and Stott (1995), Golding et al. (2003), and Wada et al. (2012), was employed. In this method, the severity of motion sickness is rated on a Likert-type scale with six levels: 1 = no symptoms, 2 = initial symptoms but no nausea, 3 = mild nausea, 4 = moderate nausea, 5 = severe nausea, and 6 = vomiting. This method is suitable for rating motion sickness in a very short time and is referred to as the *sickness rating* in the present paper. At the end of each slalom run, the experimenter sitting in the rear seat asked the participant to assess his current state in terms of the sickness rating. In addition, the participant was asked to declare if he experienced any symptom of motion sickness at any time during the run. A rating indicating motion sickness triggered the termination of that driving trial.

The symptoms of motion sickness were quantified by a motion-sickness symptom score test (Golding, Markey, and Stott 1995) twice: immediately after and 10 min after the termination of the driving test. The participant subjectively rated each symptom of motion sickness, namely, dizziness, body warmth, headache, sweating, stomach awareness, increased salivation, nausea, pallor (evaluated by the experimenter), and any additional symptoms, on four levels: 0 = none, 1 = mild, 2 = moderate, and 3 = severe. The *total symptom score* quantifying the severity of the motion sickness was calculated by summing the individual ratings (Golding, Markey, and Stott 1995).

*Apparatus*

A small passenger car with a 1000 $cm^3$ engine was used for the driving experiments. An MTi-G sensor (Xsens Technologies) was fixed to a flat place close to the shift lever of the automatic transmission to measure the 3DOF acceleration and 3DOF orientation of the vehicle. Moreover, an MTx sensor (Xsens Technologies) was attached to a cap worn by the participant to measure the 3DOF acceleration and 3DOF orientation of his head. Both sensors were connected to a laptop PC in the rear seat of the vehicle to synchronise the sensor data. The sampling time for the two sensors was 10 ms.

*Procedure*

As mentioned before, the participants gave written informed consent before the experiments. In addition, before the measurement trials, each participant attended practice trials to get accustomed to the two head-tilt conditions. During the practice trial, a device that allowed the participant to know the correct head-tilt angle when his head touched it was attached to the headrest, as shown in Figure 2 [Figure 2 near here]. Then, after 30 min of rest, the measurement trials started. The participant was seated in the navigator seat of the passenger car in a normal sitting position with a safety belt and exposed to the lateral acceleration due to the slalom driving. The experimental course was a quasi-oval track having a total length of approximately 400 m with straight parts of approximately 150 m length and curved parts of 6 m and 10 m radii (Figure 3). Eight pylons were located at 15 m intervals in each straight segment [Figure 3 near here]. The driver drove continuously at approximately 30 km/h through the pylon slalom (i.e. zigzagging to the left and right of the pylons) in the straight segment and at approximately 20 km/h in the curved segments. The drivers were instructed to maintain the velocity by using the acceleration pedal as smoothly as possible and using the brake pedal as sparingly as possible to avoid large longitudinal acceleration. They were also instructed to decelerate as smoothly as possible after slaloming to enter the curved segments and accelerate as smoothly as possible when exiting the curve segments. The drivers were aware of the head-tilt and eyes-open/closed conditions.

At the end of each slalom run, the experimenter asked the participant to indicate the severity of his motion sickness in terms of the sickness rating (Golding, Markey, and Stott 1995; Golding et al. 2003). In addition, at any time during the run, if the sickness rating reached 2 or more, the participants were asked to inform this to the experimenter. In such a case, the driving trial was terminated after the next two slalom runs. Note that the participants were told again at that time that they could stop immediately without the two additional slalom runs, but no participant rejected the additional runs. Each driving trial was terminated after 30 slalom runs (15 laps), even if the sickness rating did not reach 2 or more. The time of the termination, i.e. when a sickness rating of 2 or more was reached or when 30 slalom runs were completed, is called the driving endpoint.

**Results**

*Resultant vehicle and head motions*

The frequencies of lateral oscillations, the root mean square (RMS) vehicle lateral accelerations, and the durations of the slalom runs were compared to verify the uniformity of stimulus among the four conditions (Table 1) [Table 1 near here]. The mean frequencies were in the most provocative range (around 0.25 Hz) for motion sickness (Golding, Mueller, and Gresty 2001). The vehicle acceleration signals during a part of one experiment day were misrecorded, causing a lack of data for one day for five participants. This included two data points for centrifugal eyes-open, two for centrifugal eyes-closed, and one for centripetal eyes-open conditions. Thus, the statistical test for the RMS vehicle acceleration was conducted as two separate one-factor analyses for the head-tilt and eyes-open/closed conditions. Through this processing, a one-way repeated-measures analysis of variance (ANOVA) for the RMS vehicle lateral acceleration revealed no significant effect of the head-tilt condition ($F(1,7) = 0.039, p = 0.849$). A one-way repeated-measures ANOVA for the RMS vehicle lateral acceleration also revealed no significant effect of the eyes-open/closed condition ($F(1,7) = 0.540, p = 0.486$). Furthermore, a two-way repeated-measures ANOVA for the duration of the slalom run revealed no effects of the head-tilt ($F(1,7) = 0.888, p = 0.377$) and eyes-opened/closed ($F(1,7) = 1.732, p = 0.23$) conditions and no interaction ($F(1,7) = 0.031, p = 0.865$). A two-way repeated-measures ANOVA for the RMS head roll angle also revealed no effects of the head-tilt ($F(1,7) = 1.584, p = 0.249$) and eyes-open/closed ($F(1,7) = 0.134, p = 0.725$) conditions and no interaction ($F(1,7) = 0.001, p = 0.977$). Hence, these analyses show no evidence of significant differences in the driving conditions of the vehicle and the resultant head roll angles of the participants among the four experimental conditions.

The correlation coefficients between the vehicle lateral acceleration and the head roll angle throughout the experiment were analysed for the participants. A positive correlation coefficient would imply that the head roll motion was synchronised with the vehicle lateral acceleration in the centrifugal direction. The means and standard deviations of the correlation coefficients for all the participants were calculated. In the centripetal condition, the correlation coefficients between the vehicle lateral acceleration and the head roll were 0.61 (SD 0.14) for eyes-open and 0.48 (SD 0.13) for eyes-closed conditions. In the centrifugal condition, the correlation coefficients were -0.72 (SD 0.14) for eyes-open and -0.58 (SD 0.29) for eyes-closed conditions. From these results, it can be concluded that the head tilts associated with the vehicle lateral acceleration were controlled well.

### *Number of participants experiencing each sickness rating level at driving endpoint*

Table 2 illustrates the number of subjects who experienced each sickness rating level at the driving endpoint. No participant experienced levels 4–6 in the experiments. From the table, it can be seen that the number of participants who experienced higher sickness ratings was more in the eyes-closed condition than in the eyes-open condition. It can also be observed that the results of both the head-tilt conditions were the same in the eyes-closed condition at the driving endpoint [Table 2 near here].

### *Time course of sickness rating*

Figure 4 shows the mean sickness ratings at each slalom run for different head-tilt and eyes-open/closed conditions. It can be seen that the sickness ratings reached their final values earlier in the centrifugal condition than in the centripetal condition [Figure 4 near here]. To facilitate the analysis, the subjects who reached a sickness rating of 2 or more and stopped before 30 runs were assigned continuation values for the incomplete runs after the driving endpoint. To avoid the convergence in sickness ratings due to their saturation after the driving endpoint, the statistical analysis was confined to the first 25 runs and the sickness rating at each of these runs was considered as a different data point; these data points were used for the statistical analysis.

The Wilcoxon signed-rank test was used for the statistical analysis because of the abnormal distribution of data due to the saturation of the sickness rating at the driving endpoint. The test revealed the significant main effect of the centripetal/centrifugal head-tilt condition on the sickness ratings ($z = -2.60$, $p = 0.0091$, two-tailed), with the centripetal head tilt showing a lower sickness rating. The results also revealed the significant main effect of the eyes-open/closed condition on the sickness ratings ($z = -9.18$, $p < 0.0001$, two-tailed), with the eyes-open condition showing a lower sickness rating.

The simple main effects of the head-tilt condition on the sickness ratings for the eyes-open and eyes-closed conditions were then analysed. For the eyes-open condition, the Wilcoxon signed-rank test revealed significant differences in the sickness rating between the head-tilt conditions ($z = -2.50$, $p = 0.017$, two-tailed), with the centripetal head tilt showing a lower sickness rating. This result agrees with that obtained by Wada et al. (2012). For the eyes-closed condition, the Wilcoxon signed-rank test revealed marginally significant differences in the sickness rating between the head-tilt conditions ($z = -1.67$, $p = 0.096$, two-tailed), with the centripetal head tilt showing a lower sickness rating.

*Number of slalom runs to driving endpoint*

Table 3 shows the means and standard deviations of the number of slalom runs to the driving endpoint for all the participants in each condition [Table 3 near here]. The Wilcoxon signed-rank test was again employed to analyse the differences because of the abnormal distribution of data due to the termination of driving at 30 runs in some trials. The test revealed the significant main effect of the eyes-open/closed condition on the number of slalom runs ($z = -3.05$, $n = 16$, ties = 2, $p = 0.002$, two-tailed)—a larger number of runs to the endpoint was observed for the eyes-open condition—and no significant main effect of the head-tilt condition ($z = -1.622$, $n = 16$, ties = 1, $p = 0.105$, two-tailed). The Wilcoxon signed-rank test also revealed no significant simple main effect of the head-tilt condition on the number of slalom runs for the eyes-open ($z = -0.931$, $n = 8$, ties = 1, $p = 0.352$, two-tailed) and eyes-closed ($z = -1.193$, $n = 8$, $p = 0.233$, two-tailed) conditions. There was no evidence that the number of slalom runs was significantly different between the head-tilt conditions for the eyes-open and eyes-closed conditions.

*Total symptom score (TSS)*

Table 4 shows the total symptom scores at the driving endpoint and 10 min after it [Table 4 near here.]. A two-way repeated-measures ANOVA for the TSS at the driving endpoint revealed no effect of the head-tilt ($F(1,7) = 0.548$, $p = 0.483$) and eyes-open/closed ($F(1,7) = 0.368$, $p = 0.563$) conditions and no interaction ($F(1,7) = 0.800$, $p = 0.401$). In addition, a two-way repeated-measures ANOVA for the TSS 10 min after the driving endpoint revealed no effect of the head-tilt condition ($F(1,7) = 0.00$, $p = 1.00$) and no interaction ($F(1,7) = 0.636$, $p = 0.451$), but significant effect of the eyes-open/closed condition ($F(1,7) = 21.0$, $p = 0.00254$) was observed, with larger TSS in the eyes-closed condition.

**Discussion**

The significant decrease in the sickness rating due to the centripetal head tilt, compared with the centrifugal head tilt, in both the eyes-open and eyes-closed conditions (Figure 4) demonstrates the effect of head tilt on motion sickness. This result suggests that the carsickness of passengers could be reduced if they tilt their heads against the centrifugal direction, thus imitating the driver's head tilt, regardless of whether they keep their eyes open or closed. In addition, the significant decrease in the sickness rating in the eyes-open condition, compared with the eyes-closed condition, in both the head-tilt conditions indicates the benefit of keeping the eyes open in reducing motion sickness. This result is corroborated by the larger number of slalom runs to the driving endpoint in the eyes-open condition than in the eyes-closed condition.

The finding that the centripetal head tilt reduces motion sickness in the eyes-open condition agrees with the results obtained by Wada et al. (2012), who conducted slalom driving experiments similar to those conducted in the present study but limited to the eyes-open condition. This also agrees with the results of Golding et al. (2003), which showed that the severity of motion sickness in a longitudinal linear acceleration environment decreased when the head was actively aligned with the GIF for the eyes-open condition. The contribution of the present study is that it demonstrates the effect of centripetal head tilt in reducing the severity of motion sickness in a real automotive environment for both the eyes-open and eyes-closed situations. Thus, it is the first study to demonstrate that the finding of Wada et al. (2012), i.e. centripetal head tilt reduces motion sickness, is also true when there is no visual information of the upcoming road shape. There have also been investigations on the effect of visual information on motion sickness, and they suggest that a visual scene of the upcoming road reduces the severity of the motion sickness of passengers sitting naturally in the rear seat (Griffin and Newman 2004). Therefore, another contribution of the present study is that it demonstrates the effect of keeping the eyes open in reducing motion sickness with active head tilt in either direction.

We now consider factors contributing to motion sickness when riding a car. Rolnick and Lubow (1991) reviewed the factors contributing to the difference in motion sickness between drivers and passengers, namely, head movement, controllability, perceived control, visual information, predictability, and activity. Among these factors, controllability, perceived control, and activity were equivalent in the different experimental conditions considered in the present study. With regard to the controllability factor, Rolnick and Lubow (1991) showed that participants who felt themselves to be in control of a vehicle with 1DOF rotational motion around a vertical axis by using a joystick were less likely to experience motion sickness. In their study, a passive participant with no controllability over the vehicle motion was also prevented from moving his or her head voluntarily by being connected to the other (active) passenger's head via helmets. In the present study, the controllability factor was equivalent in the different conditions because the participants moved their heads in a predetermined manner. Perceived control or sense of control is a subjective psychological state by which the person can determine his or her behaviour. The factor of perceived control was also equivalent in both the head-tilt and eyes-open/closed conditions because the participants moved their heads by themselves and had been informed that they could stop the experiment at any time. The visual information differed in the eyes-open and eyes-closed conditions and was therefore an independent factor in this research. However, it was not very different in the two head-tilt conditions for each eyes-open/closed condition. In fact, the visual information was equivalent in the eyes-closed condition for both head tilts, and it was not very different in the eyes-open condition for the two head tilts because the participants sat in the navigator seat and observed similar road scenes ahead. It is thought that the predictability is composed of visual information and previous knowledge of the motion in this case. The predictability of the vehicle motion would be easy to acquire because the participants had been informed that they would ride along a predetermined test track and were also notified at the entrance of each slalom run that the run was beginning. However, the predictability factor differed between the eyes-open and eyes-closed conditions because of the difference in visual information, whereas it was almost equivalent between the head-tilt conditions because of the similar visual information. Activity is also thought to be a factor contributing to motion sickness, as Wendt (1951) postulated, but Rolnick and Lubow (1991) pointed out that very few studies have directly investigated the effect of this factor and one study demonstrated contradictory results. The effect of activity on motion sickness can also be understood in terms of the sensory rearrangement theory (Reason and Brand 1975) as follows: According to the reafference principle (von Holst 1954), our active movement prompts neural adaptation by comparing the efference copy of the motor command with the reafference from the effector. For example, Held and Bossom (1961) showed that prism adaptation is significantly faster in self-produced motion. Thus, according to the sensory rearrangement theory, it is understood that higher activity can reduce motion sickness because of its faster adaptation. Wada et al. (2012) compared the effect of the active head tilt towards the centripetal direction with that of the natural or passive head movement of passengers. Thus, the study had the limitation that the head movements were not equivalent between the two head-tilt conditions. In the present study, the activity of the head movements was carefully equalised by instructing the subjects to tilt their head actively. Moreover, the participants practiced tilting the head to the correct angle in advance by using the device shown in Figure 2, and there was no evidence that the head movements differed between the head-tilt conditions. The results of the present study, in which the head-tilt strategy for both the eyes-open and eyes-closed conditions had no significant simple main effect on the number of runs to the driving endpoint, did not agree with the results of Wada et al. (2012), in which a significant effect of the head tilt was observed for the eyes-open condition. It is thought that this dissimilarity may be due to the smaller difference in the activity of the head tilt in the present study.

The present study also demonstrated that the head-tilt strategy has a significant effect on the time course of the severity of motion sickness but not on the number of runs. Thus, the results of Wada et al. (2012), in which a significantly larger number of laps were considered for the centripetal head-tilt condition, can be understood as the effect of the combination of head tilt and some differences in the head-tilt activity. This implies that the motion sickness with active head tilt towards the centrifugal direction might be less severe than that with passive or natural head movement towards the same direction, which occurs during driving. Moreover, Golding et al. (2003) investigated the motion sickness due to longitudinal oscillations in an experimental cabin without any external view and for a longer duration than that in the present study. They observed that the time required to attain a sickness rating of 4 (moderate nausea) with the head motion aligned with the GIF direction was larger than that with the misaligned head motion when both head movements were actively produced, whereas the times required to attain sickness ratings of 2 and 3 did not significantly differ. These results agree with the findings of the present study. Note that Golding et al. (2003) also conducted an experiment in which the head was aligned with the GIF and misaligned by the automatic controlled suspension system in a longitudinal linear acceleration environment. The results demonstrated that motion sickness was more severe in the aligned condition than in the misaligned condition, which interestingly showed the opposite tendency to the active head-tilt condition. The result that GIF alignment by passive head motion increases motion sickness agrees with the fact that in a high-speed train with a tilting mechanism towards the GIF direction, motion sickness increases (Persson 2008).

In summary, the discussion above suggests that passengers' active head tilt towards the centripetal direction by itself reduces the severity of motion sickness in a lateral acceleration environment, regardless of the eyes-open/closed condition, and this was verified under the conditions of equivalent controllability, perceived control, visual information, predictability, and activity in the present study. On the other hand, differences were found in the number of participants who experienced a sickness rating of 2 and in the number of slalom runs at the driving endpoint only in the eyes-open/closed condition but not in the head-tilt condition. This result indicated that the opening of the eyes had a larger effect on motion sickness than the head-tilt strategy. Consequently, these results imply that both the eyes-open condition and the head tilt towards the GIF reduce motion sickness, and it is thought that their interaction is not large, but the effect of the eye opening is larger than that of the head tilt.

The SVC theory proposed by Bles et al. (1998) postulates that motion sickness is caused by the accumulation of discrepancies between the vertical direction sensed by sensory organs, such as the eyes, vestibular system, and non-vestibular proprioceptors, and the subjective vertical direction estimated from the internal model in the central nervous system. Mathematical models of the SVC theory that consider only the vestibular system have been proposed for 1DOF motion in the vertical direction (Bos and Bles 1998) and for 6DOF motion, called the 6DOF-SVC model (Wada et al. 2010), which expands the original 1DOF version and includes head rotation. By using the 6DOF-SVC model, Wada et al. (2010) and Wada, Kamiji, and Doi (2013) showed that the motion sickness incidence with the head tilted towards the centripetal direction is less than with the head kept upright or tilted towards the centrifugal direction. The lower sickness ratings in the experimental results for centripetal head tilt in the eyes-closed condition agree with the calculation results from the 6DOF-SVC model. Thus, it is hypothesised that the effect of the head tilt observed in the present research can be partially interpreted by the SVC theory. However, the effect of the visual condition on the sickness rating, which was not included in the mathematical model, was larger than that of the head-tilt condition in the experiments. This could be understood as the effect of the passenger being able to predict the upcoming road environment.

No significant differences between the head-tilt conditions were observed in the TSS at the endpoint and 10 min after the endpoint. Even though this does not agree with the results of Wada et al. (2012), it can be attributed to the fact that the endpoint was determined by the sickness rating in this study. It is thought that in Wada et al. (2012)'s study, significant differences were observed because a larger number of participants reached the driving cutoff without any symptoms in the active head movement condition. The variations in the activity of the head tilt could be another reason for the above-mentioned differences. Furthermore, the TSS 10 min after the driving endpoint in the eyes-open condition was smaller than that in the eyes-closed condition. This agrees with the effect of visual information as described above.

**Conclusion**

An experiment was conducted to examine the hypothesis that a passenger's active head-tilt motion in the lateral direction affects the severity of motion sickness in an automobile under lateral acceleration, regardless of whether the eyes are kept open or closed. The results showed that for both eyes-open and eyes-closed conditions, the mean sickness rating in the time course of the slalom driving was significantly lower in the centripetal head-tilt condition, which is similar to the drivers' head motion but opposite to normal passive passengers' head motion. It was also revealed that the time course of the sickness rating was also significantly lower in the eyes-open condition for both the head-tilt conditions. These results strongly indicate that a passenger's active head tilt towards the centripetal direction reduces the severity of motion sickness in the lateral acceleration environment of an automobile. In addition, they demonstrate that the head-tilt strategy for reducing motion sickness is valid, regardless of whether the eyes are kept open or closed. The results also demonstrate that visual information has a positive effect in reducing motion sickness under the active head-tilt strategy. Moreover, the effect of the eyes-open/closed condition is greater than that of the head-tilt strategy.

Furthermore, the number of slalom runs to the driving endpoint for the eyes-open condition was significantly greater than for the eyes-closed condition, whereas the effect of the head tilt was not statistically significant ($p = 0.105$). Both the head-tilt and eyes-open/closed conditions had no significant effect on the total symptom score. By comparing the present results with the finding of Wada et al. (2012) that the number of laps is significantly smaller with the active head tilt in the centripetal direction, it was deduced that the severity of motion sickness with the active centripetal head tilt was less than that with the passive or natural head movement, which can be seen in a natural driving environment.

In future studies, it is important to investigate whether the conclusions are valid for females and for a wide range of age groups, as the present research considered only young males. In addition, the effects of various head motions on motion sickness, including the phase difference between the acceleration and the head motion, should be studied in detail. Such studies could lead to the development of a methodology for reducing motion sickness.

Table 1. Mean (SD) of resultant vehicle and head motions

|  | Eyes open | | Eyes closed | |
|---|---|---|---|---|
|  | Centripetal head tilt | Centrifugal head tilt | Centripetal head tilt | Centrifugal head tilt |
| Frequency of lateral oscillation [Hz] | 0.26 (0.02) | 0.25 (0.02) | 0.24 (0.01) | 0.24 (0.01) |
| RMS vehicle lateral acceleration [$m/s^2$] | 2.02 (0.30) | 2.01 (0.22) | 1.90 (0.18) | 1.96 (0.21) |
| Duration of slalom run [s] | 30.7 (1.4) | 30.6 (1.1) | 31.3 (1.7) | 31.0 (0.7) |
| RMS head roll angle [°] | 12.0 (2.0) | 10.6 (3.5) | 12.2 (3.6) | 10.8 (4.0) |

Table 2. Number of participants experiencing each sickness rating level at driving endpoint

| Sickness rating | Eyes open | | Eyes closed | |
|---|---|---|---|---|
|  | Centripetal head tilt | Centrifugal head tilt | Centripetal head tilt | Centrifugal head tilt |
| 1: No symptom | 3 | 1 | 0 | 0 |
| 2: Initial symptom | 4 | 6 | 5 | 5 |
| 3: Mild nausea | 1 | 1 | 3 | 3 |
| 4: Moderate nausea | 0 | 0 | 0 | 0 |

Table 3. Mean (SD) of number of slalom runs to driving endpoint

| Eyes open | | Eyes closed | |
|---|---|---|---|
| Centripetal head tilt | Centrifugal head tilt | Centripetal head tilt | Centrifugal head tilt |
| 21.8 (7.9) | 19.4 (5.1) | 16.4 (5.5) | 13.5 (3.0) |

Table 4. Mean (SD) of TSS at driving endpoint and 10 min after driving endpoint

|  | Eyes open | | Eyes closed | |
| --- | --- | --- | --- | --- |
| TSS | Centripetal head tilt | Centrifugal head tilt | Centripetal head tilt | Centrifugal head tilt |
| At driving endpoint | 4.5 (3.9) | 5.4 (5.0) | 5.4 (3.5) | 5.3 (3.2) |
| 10 min after driving endpoint | 1.1 (1.4) | 1.5 (1.9) | 1.9 (2.5) | 1.5 (1.4) |

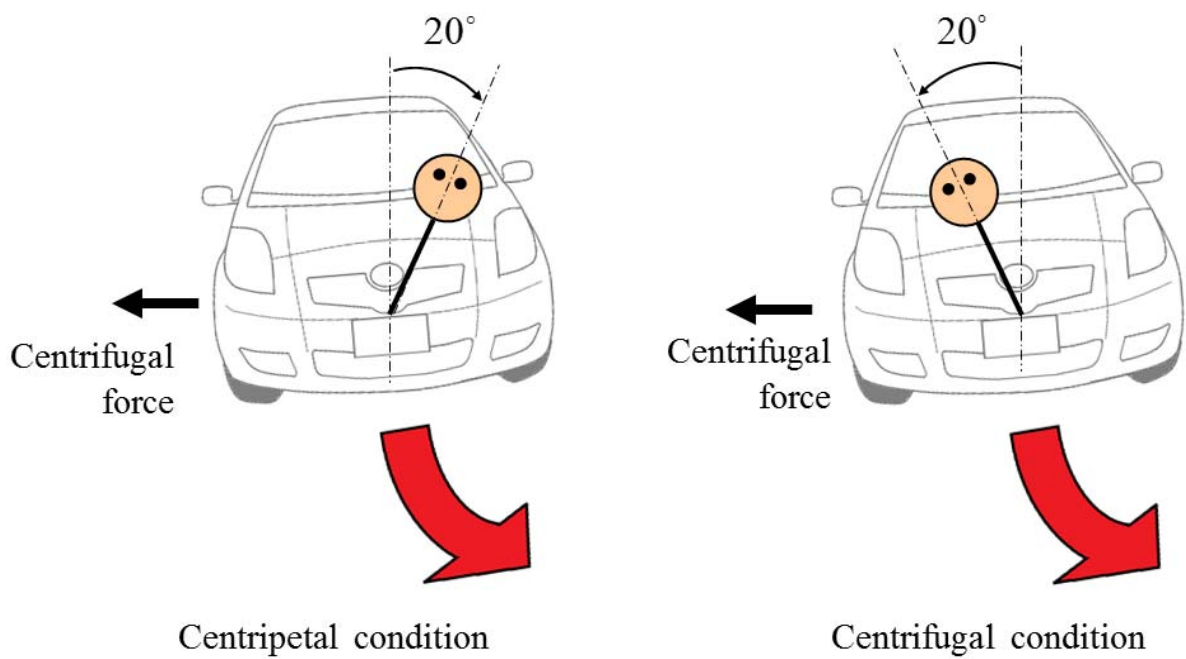

Figure 1. Typical head postures in centripetal and centrifugal conditions

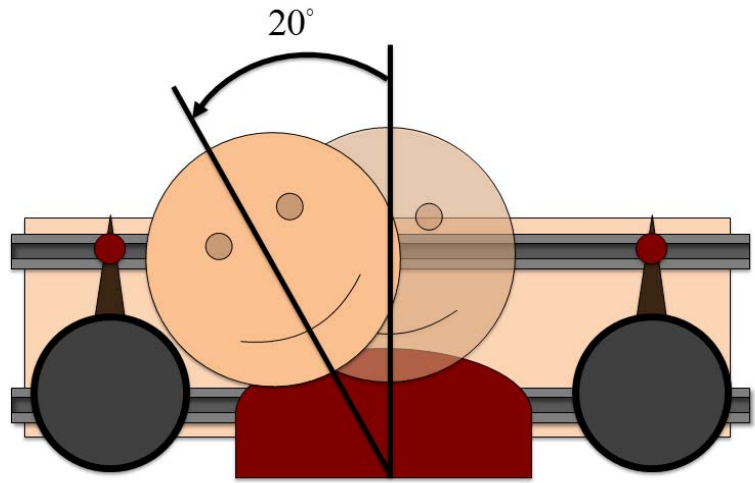

Figure 2. Device for head-tilt practice

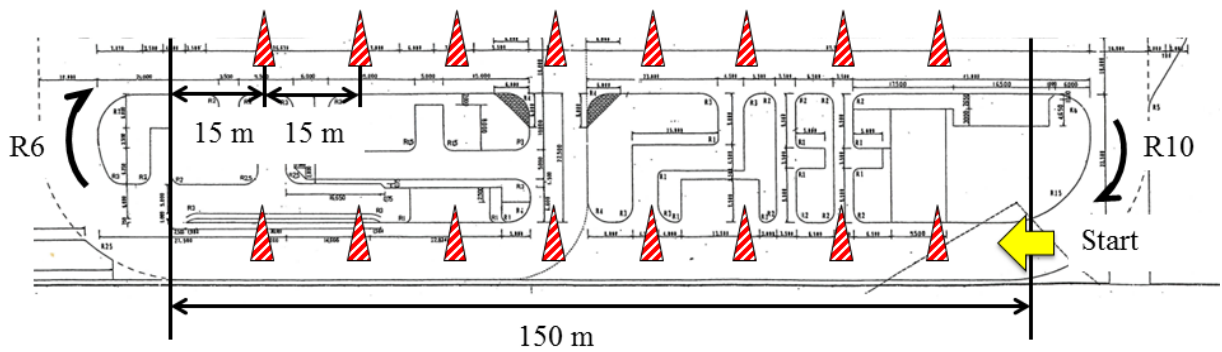

Figure 3. Test track. There are 16 pylons in slalom driving, eight for each straight segment. Zigzag driving through eight pylons on the straight segment was called a slalom run, and each lap was composed of two slalom runs.

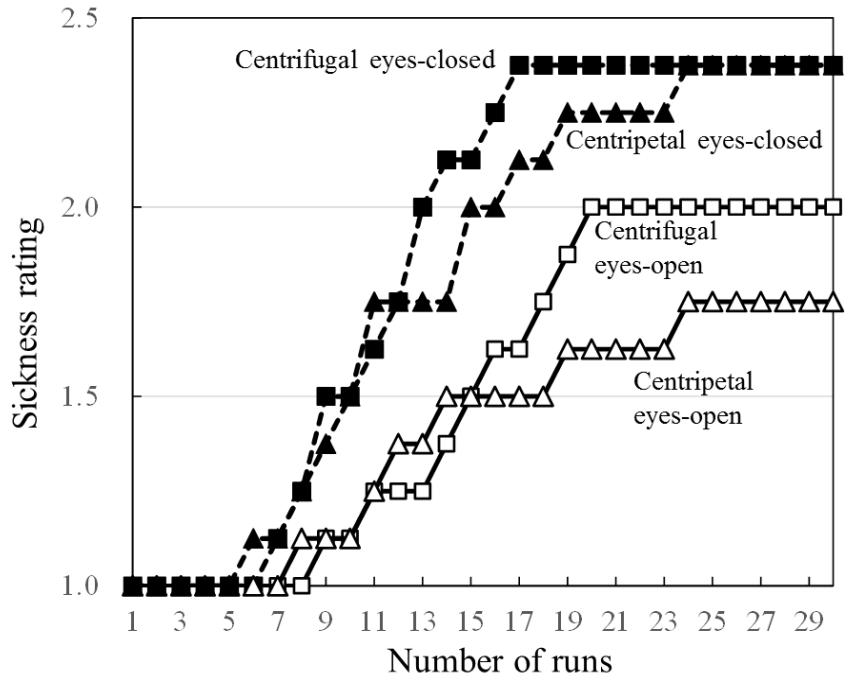

Figure 4. Mean sickness ratings at each slalom run for different head-tilt and eyes-open/closed conditions. Subjects who reached the driving endpoint with a sickness rating of 2 or more before 30 runs were assigned continuation values after the driving endpoint for the purpose of this illustration and analysis.